\documentclass[aps,prl,twocolumn,superscriptaddress]{revtex4}
\usepackage{amsmath}
\usepackage{amssymb}
\usepackage{graphicx}
\usepackage{dcolumn}
\usepackage{bm}
\makeatletter
\def\captionof#1#2{{\def\@captype{#1}#2}}
\makeatother

\begin{document}

\title{Out-of-equilibrium bosons on a one-dimensional optical random lattice}

\author{Thierry Platini}
\email{thipla@vt.edu}
\affiliation{Laboratoire de Physique des Mat\'eriaux, UMR CNRS 7556,
Universit\'e Henri Poincar\'e, Nancy 1,
B.P. 239, F-54506 Vandoeuvre les Nancy Cedex, France}
\affiliation{Department of Physics, Virginia Tech, Blacksburg, VA 24061, USA}
\author{Rosemary J. Harris}
\email{rosemary.harris@qmul.ac.uk}
\affiliation{School of Mathematical Sciences, Queen Mary, University of London, Mile End Road, London, E1 4NS, UK}
\author{Dragi Karevski}
\email{karevski@lpm.u-nancy.fr}
\affiliation{Laboratoire de Physique des Mat\'eriaux, UMR CNRS 7556,
Universit\'e Henri Poincar\'e, Nancy 1,
B.P. 239, F-54506 Vandoeuvre les Nancy Cedex, France}

\begin{abstract}
We study the transport properties of a one-dimensional hard-core boson lattice gas coupled to two particle reservoirs at different chemical potentials generating a current flow through the system. In particular, the influence of random fluctuations of the underlying lattice on the stationary state properties is investigated. We show analytically that the steady-state density presents a linear profile. The local steady-state current obeys the Fourier law $j=-\kappa(\tau)\nabla \rho $ where $\tau$ is a typical timescale of the lattice fluctuations and $\nabla \rho$ the density gradient imposed 
by the reservoirs. 
\end{abstract}

\pacs{Valid PACS appear here}
\maketitle

Ultracold atoms/molecules trapped in optical lattices have become a very active field of both experimental and theoretical research. For example, optical lattices can now be used to experimentally generate one-dimensional (1D) bosonic systems~\cite{moritz,tolra,kinoshita1} which have been studied theoretically for many years~\cite{girardeau,liebliniger,lenard}. In the large scattering-length limit and at low densities, ultracold bosons effectively behave as impenetrable particles~\cite{olshanii}, namely as hard-core bosons, thus realising the Tonks-Girardeau model~\cite{girardeau,liebliniger}. Experiments on such 1D hard-core bosons have been performed with Rubidium atoms within both continuum~\cite{kinoshita1} and lattice contexts~\cite{paredes}. 
In particular, theoretical and exprimental considerations~\cite{kinoshita2,rigol1} show
that a harmonically-trapped hard-core boson gas does not relax towards an equilibrium state even if the momentum distribution of an expanding-gas state~\cite{karevski1}  approaches that of non-interacting fermions~\cite{rigol2}. Also worthy of mention is work on the influence of finite currents on the superfluid to Mott-insulator transition in a Bose-Hubbard model~\cite{polkovnikov}.  For a recent review of developments in ultracold gases and optical lattices, see~\cite{bloch}.
In this letter we present the transport properties of a 1D-lattice hard-core bosonic gas driven out of equilibrium by the interaction at its boundaries with two external reservoirs that induce a particle-current flow through the system. We focus particularly on the influence of lattice fluctuations, which may be induced artificially or inherent to an experimental set-up, on the transport properties.

The Hamiltonian of 1D hard-core bosons is given by
\begin{equation} 
H_S=-\sum_{l=1}^{N-1} t_l[b^+_lb_{l+1}+b^+_{l+1}b_l]+\epsilon \sum_{l=1}^N n_l
\label{eq1}
\end{equation}
where $N$ is the number of lattice sites, $b^+_l$, $b_l$ are the bosonic creation and annihilation operators at site $l$, $n_l=b^+_lb_l$ is the particle number operator, $\epsilon$ the chemical potential, and $t_l$ the particle hopping rate across the $l$th bond. The creation and annihilation operators satisfy the usual bosonic commutation relations on different sites, $[b_l,b^+_{l'}]=[b_l,b_{l'}]=[b^+_l,b^+_{l'}]=0$, while the hard-core constraint is implemented by the additional conditions  $b_l^2=(b_l^+)^2=0$ and $\{b_l,b_l^+\}=1$ preventing more than single occupancy of sites.  Through the transformation $b^+=(\sigma^x+i \sigma^y)/2$, where $\sigma^{x,y}$ are the usual Pauli matrices, the hard-core boson Hamiltonian is exactly mapped onto the $XX$ quantum spin chain $H^{XX}=-1/2\sum_l t_l [\sigma_l^x\sigma_{l+1}^x+\sigma_l^y\sigma_{l+1}^y]+\epsilon/2\sum_l \sigma^z_l+const.$
 which, in recent years, has been studied extensively in a nonequilibrium context~\cite{karevski2}.

Our system is coupled to a quantum environment consisting of two independent reservoirs (at the left and the right of the system). The system-reservoir interaction is  implemented via a repeated interaction scheme described as follows \cite{AttalPautrat}.  At a given time a left(right) reservoir ``particle'' comes close to the left(right) boundary, interacts locally with the system during a short time interval $\tau$ and goes away for ever, being replaced by another fresh reservoir particle. The process is repeated again and again.  Let ${\cal H}_S$ be the system Hilbert space, ${\cal H}_n$ the Hilbert space of the $n$th environment copy with associated Hamiltonian $H_n$, and $H_E=\sum_{\mathbb{N}^*}H_n$ the total environment Hamiltonian defined on ${\cal E}=\otimes_{\mathbb{N}^*}{\cal H}_n$. We assume that initially, at $t\leq0$, the system and the environment are uncoupled and described by the factorized state $\rho(0)=\rho_S(0)\otimes\rho_E(0)$ where $\rho_S$ and $\rho_E$ are respectively the system and reservoir density matrices. By assumption, the reservoir state is factorized into $\rho_E(0)=\otimes_{\mathbb{N}^*}\rho_n$, where $\rho_n$ is the initial state of the $n$th environment subsystem.
At $t>0$ the interaction between the system and the reservoirs is switched on and the total Hamiltonian on the time interval $](n-1)\tau,n\tau]$ is given by $H_{Tot}^{\{n\}}=H_S+H_E+H_I^{\{n\}}$ where 
$H_I^{\{n\}}$ is the interaction Hamiltonian between the $n$th copy of the environment and the system. The time evolution on that interval is $K_n(\tau)=U_{I}^{\{n\}}\otimes\prod_{\mathbb{N}^*\setminus\{n\}}U_k(\tau)$
with the unitary evolution $U_k(\tau)=e^{-i\tau H_k}$ and $U_{I}^{\{n\}}=e^{-i\tau (H_S+H_I^{\{n\}}+H_n)}$.
The dynamics from $t=0$ to $t=n\tau$ is generated by the string operator
$
{\cal U}(n\tau)=K_n(\tau)K_{n-1}(\tau)\dots K_1(\tau)\; ,
$
and in the Schr\"odinger picture the total state evolves as 
\begin{equation}
\rho(n\tau)={\cal U}(n\tau)\rho(0){\cal U}^\dagger(n\tau)=K_n(\tau)\rho((n-1)\tau)K_n^\dagger(\tau)\; .
\end{equation}
The reduced density matrix associated with the system itself, $\rho_S(t)=Tr_E\{\rho(t)\}$, evolves as
$
\rho_S(n\tau)=Tr_n\left\{U_{I}^{\{n\}} [\rho_s((n-1)\tau)\otimes \rho_n] {U_{I}^{\{n\}}}^\dagger\right\}
$
where $Tr_n$ stands for the partial trace over the $n$th environment subsystem.
Introducing the eigenstates of $\rho_n$, $\{\phi^{\{n\}}\}_{j=1\dots dim\{{\cal H}_n\}}$ with eigenvalues $p^{\{n\}}_j$, such that $\rho_n|\phi^{\{n\}}_j\rangle=p^{\{n\}}_j|\phi^{\{n\}}_j\rangle$, and the decomposition $\{|\psi^S_k\rangle\otimes|\phi^{\{n\}}_j\rangle\}$ of the joined $n$th system and subsystem Hilbert space ${\cal H}_S\otimes{\cal H}_n$, we arrive at the closed dynamics
$
\rho_S(n\tau)={\cal K}_n\left(\rho_S((n-1)\tau)\right)
$
for the system.  Here ${\cal K}_n$ is the completely positive map
$
{\cal K}_n X\equiv\sum_{l,m=1}^{dim\left\{{\cal H}_n\right\}}p^{\{n\}}_l V_m^l X {V_m^l}^\dagger
$
where the coefficients $V_m^l$, which are operators on ${\cal H}_S$, are defined through the decomposition of $U_I^{\{n\}}$ on $\{|\psi^S_k\rangle\otimes|\phi^{\{n\}}_j\rangle\}$:
\begin{equation}
U_I^{\{n\}}=\left(\begin{array}{cccc}
V_1^1&V_1^2&\dots&V_1^{dim\{{\cal H}_n\}}\\
V_2^1&V_2^2&\dots&V_2^{dim\{{\cal H}_n\}}\\
\vdots&\vdots&&\vdots\\
V_{dim\{{\cal H}_n\}}^1&V_{dim\{{\cal H}_n\}}^2&\dots&V_{dim\{{\cal H}_n\}}^{dim\{{\cal H}_n\}}
\end{array}\right).
\end{equation}
Note that, for typographical convenience, we have suppressed here the dependence of the operators $V$ on $n$.
For identical subsystems, with $\rho_n=\rho_{e}$ $\forall n$, we have ${\cal K}_n={\cal K}$ $\forall n$. Iterating the previous recursive equation, we arrive at
$
\rho_S(n\tau)={\cal K}^n\left(\rho_S(0) \right).
$
The Heisenberg equation of motion is obtained through the adjoint action of ${\cal K}$ defined through the trace-like scalar product $(X,Y)\equiv Tr\{XY^\dagger\}$ via $(X,{\cal K}Y)=({\cal K}^\dagger X,Y)$.  Finally, for the time evolution of a system observable one has
$
X(n\tau)={{\cal K}^\dagger}^n X(0)\; .
$

In our setup, the interaction with the reservoirs is given by the local left- and right-hopping Hamiltonians
$H_L=-t_L[b^+_Lb_{1}+b^+_{1}b_L]$, $H_R=-t_R[b^+_Rb_{N-1}+b^+_{N-1}b_R]$, describing the exchange of particles through the boundaries between the reservoirs and the system. In the following we set $t_L=t_R=t_E$.
The left and right reservoir particles are described by the one-particle density matrices
$\rho_{L,R}=\left(z_{L,R}\right)^{n_{L,R}}=|1\rangle p_1^{L,R}\langle 1|+|0\rangle p_0^{L,R}\langle 0|$
where $z_{L,R}$ are the left and right fugacities, $|0\rangle,|1\rangle$ the vacuum and one particle states and $p_0$, $p_1$ their respective occupation probabilities, with $p_1+p_0=1$.  

The hard-core boson Hamiltonian (\ref{eq1}) is canonically diagonalized by introducing a fermionic representation of the creation and annihilation operators $b^+$, $b$~\cite{lieb}:
\begin{eqnarray}
{\Gamma_S}_l^1=e^{i\pi\sum_{j=1}^{l-1}n_j}(b_l+b_l^+)\\
{\Gamma_S}_l^2=ie^{i\pi\sum_{j=1}^{l-1}n_j}(b_l-b_l^+)
\end{eqnarray}
where the $\Gamma_S$s are Majorana real (Clifford) operators satisfying ${\Gamma_S}^\dagger=\Gamma_S$ and $\{{\Gamma_S}_i^\alpha,{\Gamma_S}_j^\beta\}=2\delta_{ij}\delta_{\alpha\beta}$. In the Clifford representation, the Hamiltonian (\ref{eq1}) takes the form 
$
H_S=(1/4) {{\bf \Gamma}_S}^\dagger T_S {\bf \Gamma}_S
$
where ${{\bf \Gamma}_S}^\dagger =({\Gamma_S}_1^1,{\Gamma_S}_2^1,...,{\Gamma_S}_N^1,{\Gamma_S}_1^2,...,{\Gamma_S}_N^2)$
is a $2N$-component operator and 
$
T_S=\left(\begin{array}{cc}
0&C_S\\
C_S^\dagger&0
\end{array}
\right)
$
is a $2N\times 2N$ matrix written in terms of the tridiagonal matrix $[C_S]_{jk}=-i[\epsilon \delta_{jk}+t_j(\delta_{jk+1}+\delta_{jk-1})]$ with boundary conditions $t_0=t_{N}=0$.
The time evolution of ${\bf \Gamma}_S$, generated by $H_S$, is simply given by ${\bf \Gamma}_S(t)=e^{-itT_S}{\bf \Gamma}_S(0)\equiv {\bf R}_S(t){\bf \Gamma}_S(0)$, which defines the rotation matrix ${\bf R}_S$. Its matrix elements are expressed in terms of the spectral properties of $H_S$, see~\cite{karevski1} for the explicit forms.

In the repeated interaction procedure, where the $n$th copy of the environment interacts with the system on time-interval $](n-1)\tau,n\tau]$, the interaction Hamiltonian $H_S+H_I^{\{n\}}+H_n$ is again of the form (\ref{eq1}) with one more site, either on the left of the system if it is a left reservoir particle that interacts or on the right if it is a right reservoir particle. We introduce correspondingly a $2(N+1)$-component Clifford vector operator ${\bf \Gamma}^{\{n\}}$ whose dynamics is governed by the new $2(N+1)\times 2(N+1)$ interaction matrix $T^{\{n\}}$ via ${\bf \Gamma}^{\{n\}}(\tau)=e^{-i\tau T^{\{n\}}}{\bf \Gamma}^{\{n\}}$. Exactly at time $n\tau$ the expectation value of an observable
$\bf Q$, which has a given decomposition ${\bf Q}=F_Q({\bf \Gamma}^{\{n\}})$ on the $\Gamma^{\{n\}}$s, is given by
$
\langle {\bf Q}\rangle(n\tau)=Tr_{S,n}\left\{F_Q({\bf \Gamma}^{\{n\}}) U_{I}^{\{n\}} [\rho_s((n-1)\tau)\otimes \rho_n] {U_{I}^{\{n\}}}^\dagger \right\}=Tr_{S,n}\left\{F_Q({\bf \Gamma}^{\{n\}}(\tau))[\rho_s((n-1)\tau)\otimes \rho_n]\right\}\; .
$
Due to the initial Gaussian factorized state describing the system and the environment, the reduced density matrix associated with the system itself remains a Gaussian state during the repeated interaction process. One has 
$
\rho_S(n\tau)=\frac{1}{Z_S(n\tau)} e^{-\frac{1}{4}{{\bf \Gamma}_S}^\dagger T_S(n\tau){\bf \Gamma}_S}
$
where $T_S(n\tau)$ is the time-evolved system interaction matrix. This implies that one may completely characterize the system by its two-point correlation matrix $G_S=\langle -i{{\bf \Gamma}_S}{\bf \Gamma}_S\rangle+i$, where we have subtracted the trivial diagonal part by adding $i$ to the definition. In the same way, the 
correlation matrix $G^{\{n\}}=\langle -i{\bf \Gamma}^{\{n\}}{\bf \Gamma}^{\{n\}}\rangle+i$  completely characterizes the state of the system interacting with the $n$th environment copy. 
Using the somewhat shortened notation $\rho_{Sn}$ for $\rho_S((n-1)\tau)\otimes \rho_n$,
one has the dynamical equation
$
G^{\{n\}}(n\tau)=R^{\{n\}}(\tau)G^{\{n\}}((n-1)\tau)R^{\{n\}}(\tau)^\dagger
$
with 
\begin{eqnarray}
G_{k,k'}^{\{n\}}(n\tau)=Tr_{S,n}\left\{-i\Gamma_k^{\{n\}}(\tau)\Gamma_{k'}^{\{n\}}(\tau)\rho_{Sn} \right\}+i\delta_{k,k'}\\
G_{k,k'}^{\{n\}}((n-1)\tau)=Tr_{S,n}\left\{-i\Gamma_k^{\{n\}}\Gamma_{k'}^{\{n\}} \rho_{Sn} \right\}+i\delta_{k,k'}\; ,
\end{eqnarray}
where we use the convention $\Gamma_k$ for the $k$th component of the vector $\bf \Gamma$.
The correlation matrix $G^{\{n\}}((n-1)\tau)$ gives the two-point correlation just before the $n$th interaction takes place, while the matrix $G^{\{n\}}(n\tau)$ contains the correlations after that interaction has taken place. 
Ordering the ${{\bf \Gamma}^{\{n\}}}^\dagger=({\bf \Gamma}^\dagger_E,{\bf \Gamma}^\dagger_S)$ such that the first part, ${\bf  \Gamma}_E$, is associated with the interacting part of the environment and the second part, ${\bf \Gamma}_S$, with the components of the system, one notices that the correlation matrix $G_{k,k'}^{\{n\}}((n-1)\tau)$ takes a block-diagonal form [due to the uncorrelated state $\rho_s((n-1)\tau)\otimes \rho_n$]:
\begin{eqnarray}
G^{\{n\}}((n-1)\tau)=\left(\begin{array}{cc}
G_E&0\\
0&G_S((n-1)\tau)
\end{array}\right)\; .
\end{eqnarray}
Decomposing the rotation matrix as
$
{\bf R}^{\{n\}}=\left(\begin{array}{cc}
R_E&R_{ES}\\
R_{SE}&R_S
\end{array}\right),
$
one arrives at the fundamental dynamical equation for the system correlation matrix
\begin{equation}
G_S(n\tau)={ R}_S(\tau) G_S((n-1)\tau){{ R}_S}^\dagger(\tau)+{ R}_{SE}(\tau) G_E{{ R}_{SE}}^\dagger(\tau)\; .
\label{eqfond}
\end{equation}
Here $R_S$ is a $2N\times 2N$ square matrix and $R_E$ a $2\times 2$ one, while $R_{ES}$ is a $2\times 2N$ rectangular matrix and $R_{SE}$ a $2N\times2$ one.  Notice that, for non-interacting dynamics, $R_{ES}=0$, $R_{SE}=0$ and the rotation matrix splits into a block-diagonal form where $R_S={\bf R}_S$ and $R_E$ are the rotation matrices of the system and environment part respectively.
Note also that in the dynamical equation (\ref{eqfond}), the bath properties enter only through the initial environment particle state, $G_E$, which stays constant in time since at each step of the repeated interaction procedure the bath particle is replaced by a fresh one.


In the following, we concentrate mainly on the steady-state properties of the bosonic system, in particular its density profile and particle current.  The hopping dynamics conserves particles and so one may define the current through the Heisenberg equation of motion $\dot{n}_l=i[H_N,n_l]\equiv {\cal J}_{l-1}-{\cal J}_l$ where ${\cal J}_l$ denotes the particle-current operator associated with the $l$th bond and given by ${\cal J}_l\equiv t_l { J}_k=it_l\left[b_lb_{l+1}^+-b_l^+b_{l+1}\right]$ which is easily expressed in terms of the Clifford operators $\Gamma$.

If the lattice is free of any disorder then, since the particles are identical, the hopping rates are uniform, i.e., $t_l=t_S$ $\forall l$. In this case, independent of the physical parameters $t_S/\epsilon$, $t_{E}$ and $\tau$, the steady-state density profile is flat except at the sites directly in contact with the reservoir particles. The density in the flat region is the mean value set by the reservoirs  $n^*_l=\bar{\rho}=(\rho_L+\rho_R)/2$ $\forall l\ne 1,N$~\cite{platkar} while the boundary values are 
$n^*_1=\bar{\rho}  - \Delta(t_S)  (\rho_{R}-\rho_{L}) / 2$
and
$n^*_N=\bar{\rho} + \Delta(t_S)  (\rho_{R}-\rho_{L}) / 2$ with a shift from the mean $\bar{\rho}$ depending on the strength $t_S$ and density difference $\rho_{R}-\rho_{L}$. The steady-state current $j^*$ takes a constant value 
$j^*= -\frac{ \pi^2\tau }{8t_S }( \rho_{R}-\rho_{L} ) $
independent of the system-size, which can be related to the fact that the total current commutes with the system Hamiltonian. In this case there is no finite conductivity $\kappa$ and the system obviously does not obey Fourier's law.

Next we consider the effect of fluctuations of the lattice parameters, which may well be induced artificially, on the steady-state properties. In particular, we take the limit of a strongly localized lattice gas, with $t_S/\epsilon,t_E/\epsilon\rightarrow 0$, where the effect of the fluctuations is to enhance significantly the local hopping rates. The simplest case of this scenario is to consider that only one bond is activated during a given time $\tau$ which is of the order of the interaction time with the reservoirs.  
At each timestep $\tau$, a single bond is activated at random and locally the particles are exchanged with a rate $t_S=1/2$, either within the system if the selected bond is a system one or with the reservoirs if the fluctuations act close to the boundaries. We introduce the expectation values of the density gradient on the $l$th bond: $\delta_l\equiv \langle n_{l+1}\rangle -\langle n_{l}\rangle$ and the associated current $j_l=\langle J_l\rangle$ where $\langle\; .\;\rangle\equiv Tr_{S}\{\;.\;\rho_S\}$.
In the weak-hopping limit $t_S\rightarrow 0$, the density gradient on bond $l$ is changed only if the activated bond is $l-1$, $l$ or $l+1$.
From the dynamical equation (\ref{eqfond}), if the hopping is enhanced on bond $l$ then 
$\delta_l$ is mapped to 
$
\delta_l'=\cos \tau \delta_l+\sin \tau j_l
$
and $j_l$ is mapped to
$
j_l'=-\sin \tau \delta_l+\cos\tau j_l
$
.
Alternatively, if the activated bond is $l\pm 1$, the updated density gradient satisfies
$
\delta_l'=\delta_l-\frac{1}{2}(\cos\tau\delta_{l\pm1}+\sin\tau j_{l\pm1}-\delta_{l\pm1})
$
while the updated current is given by
$
j_l'=\cos(\tau/2) j_l+P_{l\pm 1}
$
where the $P_{l\pm1}\propto \Gamma_l\Gamma_{l\pm2}$ are proportional to correlations across two bonds. 
Under this dynamics, the system relaxes exponentially toward a current-carrying nonequilibrium steady-state~\cite{platkar}.
Nevertheless, we remark here that the periodicity of the dynamics implies a strong slowing-down of the relaxation to the steady-state in the neighborhood of $\tau=n2\pi$.  For even values of $n$ the dynamical generator maps to the identity and for odd values it maps to a reflection dynamics which loses its relaxation properties;
below we avoid these special $\tau$ values.

In the steady state, the $P$ terms vanish on average and the set of dynamical equations for the gradient density and current closes. At any given time, the last update on bond $l$ has probability $1/3$ to have resulted from an update on bond $l$ and probability $1/3$(1/3) to have resulted from an update on bond $l-1$($l+1$). Consequently, the steady-state averaged (denoted by a star) gradient obeys
\begin{equation}
(\cos\tau -1)\delta^*_l+\sin\tau j^*_l=\eta(\tau) \label{ssgrad}
\end{equation}
where $\eta(\tau)$ is a constant independent of the bond index $l$. Since the steady-state current $j^*_l=j^*$ is constant in space it also follows from~\eqref{ssgrad} that the gradient density is site-independent and we can thus omit the $l$-subscripts.
The steady-state current satisfies
$
j^*=\frac{1}{3}\left[-\sin\tau \delta^*+\cos\tau j^*\right]+\frac{2}{3}\cos\frac{\tau}{2} j^*
$
which is equivalent to 
\begin{equation}
j^*=-\frac{\sin\tau}{3-\cos\tau-2\cos\frac{\tau}{2}}\delta^*\equiv - \kappa(\tau)\delta^*
\end{equation}
and defines the conductivity coefficient $\kappa(\tau)$.
We now determine the constant $\eta(\tau)$ by considering the boundary conditions. Remembering that the densities on the reservoir sites are fixed and that the boundary currents $j_0=j_N=0$ due to the repeated interaction scheme, one sees that an update on bond $0$ (between left reservoir and boundary site) gives $\delta'_0=\frac{1}{2}(1+\cos\tau)\delta_0$ whereas an update on bond $1$ gives
$\delta_0'=\delta_0-\frac{1}{2}(\cos\tau \delta_1+\sin\tau j_1-\delta_1)$.
The steady-state average  $\delta_0^*$ must therefore obey
\begin{equation}
\delta_0^*=\frac{1}{2}\left(\frac{1+\cos\tau}{2}\delta_0^*\right)+\frac{1}{2}\left(
\delta_0^*-\frac{\cos\tau \delta^*+\sin\tau j^*-\delta^*}{2}
\right)
\end{equation}
giving 
$
\eta(\tau)=(\cos \tau-1)\delta_0^*\;.
$
By symmetry, at the right-boundary we find $\delta_N=\delta_0$. Noting that the reservoir density difference $\Delta\rho\equiv \rho_R-\rho_L=2\delta_0^*+(N-1)\delta^*$, one finally gets for the bulk steady-state density gradient
\begin{equation}
\delta^*=\frac{\Delta\rho}{N+1+\gamma(\tau)}
\end{equation}
with the finite-size shift function (extrapolation length) $\gamma(\tau)=2\frac{\sin\tau}{1-\cos\tau}\kappa(\tau)$.
This analytical expression is compared with numerical simulation data in Fig.~1 and the agreement is seen to be excellent.
\begin{figure}
\centerline{
\includegraphics[width=8.5cm,angle=0]{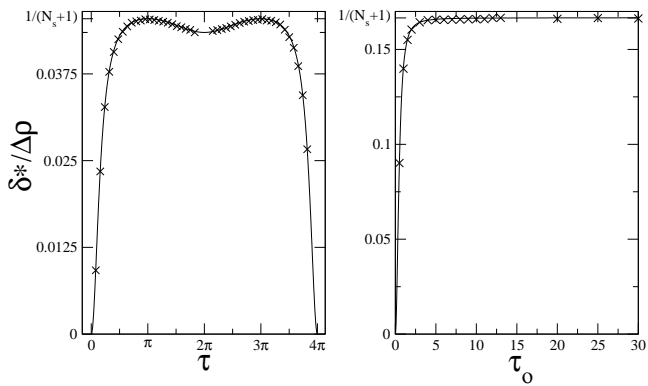}
}
\caption{\label{fig1} Normalized steady-state density gradient as a function of the interaction time with a delta time-distribution on the left and an exponential one with mean $\tau_o$ on the right. The full lines correspond to the analytical curves while the crosses are obtained numerically with a time average of the density gradient.
}
\end{figure}

So far, we have considered the somewhat unphysical situation where the enhancement of a local hopping rate always stays precisely for a time $\tau$. This hypothesis leads to the trigonometric form of the conductivity $\kappa(\tau)$ and shift function $\gamma(\tau)$.  A more reasonable assumption would be to draw the duration of a local fluctuation from a probability distribution $f(\tau)$. In that case, one may average the dynamical equations for the current and the gradient density over the time-distribution $f(\tau)$ which basically leads to replacing the trigonometric functions $\cos\tau$, $\sin\tau$ and $\cos \frac{\tau}{2}$ by their expectations under $f$. For example, for an exponential distribution of interaction timescales with mean $\tau_o$, one gets the shift function $\gamma=(2/\tau_o) \kappa$ with conductivity $\kappa=\frac{\tau_o^2+4}{3\tau_o(\tau_o^2+2)}$ which is again in agreement with the numerical results, see Fig.~1.
In the limit of short-time interaction $\tau_o\rightarrow 0$, conductivity diverges as $\tau_o^{-1}$ while the density gradient goes as $\tau_o^2$ leading to a linear vanishing of the steady-state current $j^*\sim \tau_o$.


At finite, but small, hopping rate $t_l=t_S$ along the optical lattice, we observe numerically that the linear density profile survives. However, the gradient density is strongly attenuated by a function, of the bulk hopping rate $t_S$, whose asymptotic behaviour is $(N t_S)^{-1}$ for large lattice sizes.  Consequently, for large systems the steady-state gradient density behaves as $\delta^*\sim1/N^2$ instead of the former $1/N$ behaviour. At the same time, the steady-state current $j^* \sim 1/N$, leading to a linear divergence of the conductivity coefficient $\kappa$ with system size. This implies that the classical transport properties of the hard-core boson gas are induced by the optical lattice fluctuations only for sufficiently small $Nt_S$ values.

In conclusion, we have derived analytical expressions for the steady-state conductivity and density profile of a nonequilibrium hard-core boson model. For a perfect optical lattice, due to the integrability of the model, the transport properties are anomalous with an infinite conductivity coefficient. On the other hand, when fluctuations of the underlying lattice are present and when $Nt_S\ll 1$, the classical Fourier law is recovered, with a a linear density profile and a finite conductivity coefficient.

\emph{Acknowledgements:} RJH thanks the Laboratoire de Physique des Mat\'eriaux, Universit\'e Henri Poincar\'e, Nancy for kind hospitality.  We are grateful to the Groupe de Physique Statistique there for useful discussions.

\end{document}